\documentclass[9pt,twocolumn,twoside]{opticajnl}
\journal{opticajournal} 

\setboolean{shortarticle}{true}

\usepackage{lineno}

\title{Curvature-adaptive gigapixel microscopy at submicron resolution and centimeter scale}

\author[1,$\dag$]{Xi Yang}
\author[1,$\dag$]{Haitao Chen}
\author[1]{Lucas Kreiss}
\author[1]{Clare B. Cook}
\author[1]{Genevieve Kuczewski}
\author[2]{Mark Harfouche}
\author[1]{Martin O. Bohlen}
\author[1,2,*]{Roarke Horstmeyer}

\affil[1]{Department of Biomedical Engineering, Duke University, Durham, North Carolina 27708, USA}
\affil[2]{Ramona Optics Inc., 1000 W Main St., Durham, North Carolina 27701, USA}
\affil[$\dag$]{The authors contributed equally to this work}
\affil[*]{roarke.w.horstmeyer@duke.edu}

\begin{abstract}
Large-area microscopy with submicron resolution is limited by tradeoffs between field of view (FOV), resolution, and imaging speed. Samples are rarely flat across centimeter-scale FOV, which often requires existing solutions to use mechanical scanning to ensure focused capture at reduced throughput. Here, we present PANORAMA, a single-shot, re-imaging microscope that achieves seamless, gigapixel imaging over a 16.3~$\times$~18.8~$\text{mm}^2$ FOV at 0.84~µm half-pitch resolution without mechanical scanning. By using a telecentric photolithography lens, a large-aperture tube lens, and a flat micro-camera array with adaptive per-camera focus control, PANORAMA maintains submicron focus across flat, curved or uneven samples that span centimeters. This approach improves imaging throughput and adaptability, enabling gigapixel multi-modal microscopy of large flat and non-flat samples in one shot, thus broadening its applications in biomedical and materials imaging.
\end{abstract}

\setboolean{displaycopyright}{false} 

\begin{document}

\maketitle
Large-area, high-resolution microscopy has become increasingly vital for fundamental biological research, clinical diagnostics, and industrial inspection~\cite{zhou2024computational}. However, conventional microscopes face an inherent trade-off between field of view (FOV) and spatial resolution – the number of resolvable pixels in an image (the space-bandwidth product, SBP) is fundamentally limited by objective design and sensor size. In practice, attaining gigapixel-scale images has required costly specialized optics and sensors, or time-consuming tiling and stitching of smaller high-magnification images. This bottleneck is especially problematic for emerging needs such as whole-slide histology imaging~\cite{kim2024rapid}, cortex-wide neuronal activity monitoring~\cite{fan2019video}, and imaging freely moving organisms with fine details~\cite{zhou2023parallelized}. Moreover, modern applications demand multi-modal imaging capabilities~\cite{multi_modal}, including fluorescence, phase, and polarization modes—which conventional systems struggle to accommodate at large scale.

Efforts to overcome these limitations have led to a range of innovative approaches, including scanning-based systems such as the Mesolens~\cite{mcconnell2016novel} and RA-WiFi~\cite{curve_imging}, and direct wide-field imaging platforms use CMOS sensors with a few hundred megapixels~\cite{ichimura2025volumetric}. Of note are prior "re-imaging" architectures that form a curved intermediate image with a large primary objective lens that is captured by a curved camera array~\cite{AWARE_2,fan2019video}. While such multi-scale lens systems for re-imaging inspire this current design, this prior work did not address the critical challenge of dynamically maintaining macroscopic surfaces in sharp focus.

To address this challenge, existing designs must sacrifice image throughput, reduced depth axial resolution, and/or increased system complexity to maintain focus. Alternatively, the multi-camera array microscope (MCAM) uses a densely array of micro-cameras to directly image the same sample from different, spatially offset locations in parallel to circumvent the SBP-limit~\cite{harfouche2023imaging}. MCAMs have been used for various different applications, e.g., to estimate depth through stereo imaging with overlapping views~\cite{zhou2023parallelized,zhou2024computational}, to render high-speed 3D tomography videos~\cite{zhou2025high}, to extent the depth of field (DOF)~\cite{mcam_face} or to image large histology slides at fine resolution~\cite{kim2024rapid}. Although the latter configuration can achieve microscopic resolution of down to 0.6~µm across a 54~$\times$~72~$\text{mm}^2$, it requires mechanical scanning to fill the gaps in the FOVs between adjacent cameras and thus acquire complete 2D or even 3D datasets~\cite{harfouche2023imaging}. For snapshot-based, full-FOV imaging at microscopic resolution, we previously introduced the M-FAST system, which implemented a re-imaging design with large primary lens that relayed the intermediate image plane onto a flat MCAM across a 6.6~$\times$~9.7~$\text{mm}^2$ FOV at 2.2~-~2.76~µm resolution~\cite{yang2023multi}. Although this approach effectively mitigated vignetting through the use of a fiber bundle relay, it incurred a significant loss in optical throughput~\cite{yang2023multi}.

\begin{figure*}[t]
\centering
\includegraphics[width=\textwidth]{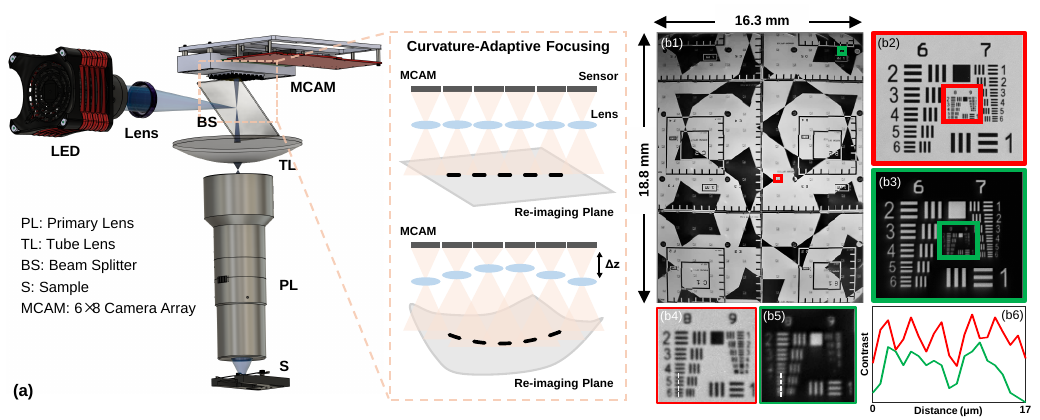}
\caption{Optical setup figure and the system resolution performance. (a) Schematic of the PANORAMA. The system is designed to relay and capture high-resolution images of the sample (S) placed at the object plane. Insert: Curvature-adaptive focusing of MCAM lenses by $\Delta z$ via per-camera focus control. (b) shows the system resolution and FOV. (b1) Full-field image of a USAF resolution target, demonstrating the large field of view (16.3~mm~$\times$~18.8~mm). (b2–b3) Magnified regions from the center (red) and corner (green) of the field, respectively. (b4–b5) Close-up views of the smallest group elements from the regions in (b2–b3). (b6) Contrast profile along a horizontal line across groups G8E5 and G8E6, comparing center (red) and corner (green) resolution performance.}
\label{fig1}
\end{figure*}

In this work, we present an improved re-imaging architecture called PANORAMA (Parallel Adaptive Non-scanning Optical Re-imaging using a Multi-camera Array) that improves optical resolution and reduces vignetting. PANORAMA is a single-shot imaging system that enables seamless, continuous coverage of a 16.3~$\times$~18.8~$\text{mm}^2$ FOV at 0.84~µm half-pitch resolution, generating 630~MP images per snapshot without scanning. Moreover, it allows for flexible focusing of each micro-camera across the re-imaged plane, which can be used to efficiently image flat, curved or uneven samples, expanding its applicability to more complex biological specimens. A detailed comparison of PANORAMA and prior systems is provided in Table~\ref{table1}. We provide a comprehensive characterization of PANORAMA’s imaging performance [Fig.~\ref{fig1}~(b) and Supplement 1 (S1)] and demonstrate its capabilities by capturing an entire histology slide in one snapshot, as well as by showcasing adaptive focus control for curved samples in both brightfield and fluorescence imaging modes.

\begin{table}[t]
\caption{\bf Comparison of PANORAMA and prior systems}
  \label{table1}
  \centering
\begin{tabular}{lcccccc}
\hline
\textbf{System} & \textbf{Res.$^{a}$ (µm)} & \textbf{FOV (mm$^2$)} & \textbf{Acq.$^{b}$} & \textbf{Curv.$^{c}$} \\
\hline
\textbf{M-FAST} & 2.2 & 6.6$\times$9.7 & SS$^{d}$ & No \\
\textbf{MCAS} & 0.6 & 54$\times$72 & TS$^{e}$ & No \\
\textbf{RUSH} & 1.2 & 10$\times$12 & SS & No \\
\textbf{AMATERAS-2} & 1.1 & 10$\times$15 & SS & No \\
\textbf{RA-WiFi} & 2.18 & 12.8$\times$12.8 & TS & Yes \\
\textbf{PANORAMA} & 0.84 & 16.3$\times$18.8 & SS & Yes \\
\hline
\end{tabular}
$^a$Res.: Resolution. \quad $^b$Acq.: Acquisition. \quad $^c$Curv.: Curvature.
$^d$SS: Single-shot. \quad $^e$TS: Tile scanning.
\end{table}

The PANORAMA extends our previous cascaded microscopy platform, M-FAST~\cite{yang2023multi}, by incorporating an improved optical design that enhances resolution and optical efficiency. The system, shown in Fig.~\ref{fig1}~(a), comprises three primary components: a high-performance primary objective lens, a large-aperture tube lens, and a flat MCAM. In the earlier M-FAST configuration, vignetting caused by the primary lens was addressed using a fiber faceplate array (FFA). While effective, it introduced a $\sim$16\% reduction in optical throughput. To overcome this limitation, PANORAMA employs a commercially available photolithography lens with image-side telecentricity to eliminate vignetting across the entire FOV of the micro-camera array. Specifically, we used the Carl Zeiss S-Planar 436~nm photolithography lens, which features an object-side numerical aperture (NA) of 0.38 and a focal length of 68~mm. This configuration offers a theoretical resolution of 0.82~µm at the 510~nm wavelength and forms a large intermediate image in its image plane spanning 16.31~mm~$\times$~18.84~mm. Following the photolithography lens, a tube lens (Ross Optical) is used to scale the intermediate image.

The PANORAMA system used the epi-illumination to provide uniform illumination for both brightfield and fluorescence imaging. A collimated, high-power LED light source (Thorlabs, SOLIS-470C) was used and coupled into the optical path through a 50/50 beam splitter or dichroic mirror (Chroma, T495lpxr). We used an array of bandpass filters (Chroma, ZET510/10x) for both imaging modalities: to reduce chromatic aberration in brightfield imaging (due to the photolithography lens design) or as an emission filter for fluorescence imaging. Thus, the fluorescence imaging resolution is effectively identical to that of brightfield.

\begin{figure}[t]
\centering
\includegraphics[width=\linewidth]{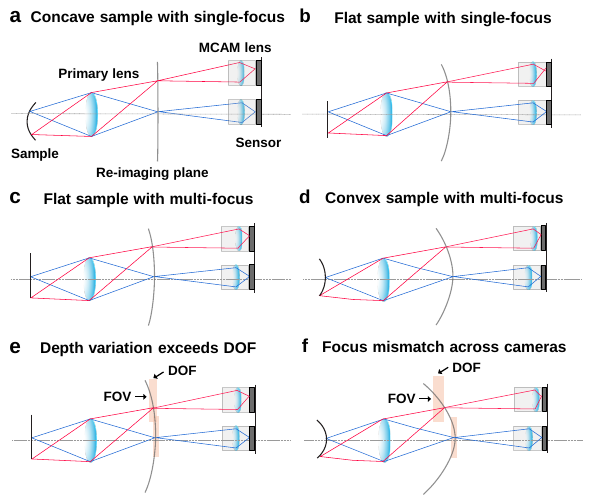}
\caption{Imaging configurations for different sample geometries and focusing strategies. (a) Concave sample with single focus yields near-flat re-imaging. (b) Flat sample with single focus shows defocus from field curvature. (c) Multi-focus corrects defocus for flat samples. (d) Multi-focus enables uniform focus for convex samples. (e) Excessive depth variation within a camera’s FOV causes local defocus. (f) Large inter-camera focus differences lead to stitching mismatch.}
\label{fig2}
\end{figure}

However, the physical dimensions of the image sensors prevent seamless tiling, which poses a challenge for large-area imaging in one shot using single-sensor architectures. In addition, field curvature grows rapidly with both field angle and NA, resulting a weakly curved spherical intermediate image plane rather than a plane~\cite{fan2019video}. Interestingly, in cases where the sample itself is concave, such as tissue sections mounted on inward-curved substrates, the sample curvature can partially counteract the optical field curvature, effectively flattening the conjugate image surface. Under such conditions, the re-imaged field approximates a plane, allowing the MCAM to operate in a default co-planar configuration [Fig.~\ref{fig2}~(a)]. In contrast, when imaging flat samples, such as standard histological slides, the optical field curvature is uncompensated, resulting in a convex intermediate image surface, producing significant defocus [Fig.~\ref{fig2}~(b)].

To efficiently capture this high-resolution, wide-area spherical image, we adopt a divide-and-conquer strategy using our MCAM. The MCAM consists of a dense array of 48 micro-cameras, each individually focused onto a local portion of the curved intermediate image plane to relay the optical signal to a flat sensor array [Fig.~\ref{fig2}~(c)]. Each unit in the MCAM utilizes a custom-designed lens with a focal length of 14.64~mm, which re-images the intermediate plane onto its respective monochrome CMOS sensor (ONSemi AR1335). The MCAM is configured as a 6~$\times$~8 array arranged in a regular grid with a center-to-center spacing of 9~mm. At this pitch and optical magnification, the FOVs of adjacent cameras overlap by approximately 10\% along the short axis and 30\% along the long axis of each sensor, which ensures complete coverage across the array while supporting accurate registration. This per-camera focus adjustment allows the system to conform to field curvature without mechanical scanning or active optical compensation, enabling uniform, in-focus gigapixel imaging over a centimeter-scale FOV.

Beyond compensating for the curvature of the intermediate image plane, each micro-camera lens is also independently focusable to match the geometry of non-planar or contoured samples [Fig.~\ref{fig2}~(d)]. This flexibility allows the effective focal surface of the entire array to be reconfigured to accommodate sample tilt, undulations, or global curvature. Note that this approach is fundamentally constrained by both the angular coverage and the DOF of each micro-camera. First, if the axial depth variation across a single micro-camera’s FOV exceeds its DOF, local defocus will occur even within that small region [Fig.~\ref{fig2}~(e)]. Second, for highly curved samples, neighboring cameras may require different focal positions. Inter-camera focus disparity introduces magnification mismatch, since axial refocusing alters the effective imaging geometry and local scaling [Fig.~\ref{fig2}~(f)]. In practice, we estimated the effective magnification of each lens based on the overlap ratios of adjacent sub-images. Then, this estimation was refined through Hugin (panorama stitcher), using pairwise geometric transformations [Supplement 1 (S2)].The magnification across the array ranged from 1.3095 to 1.4587, with adjacent camera differences up to $\sim$10\%; empirically, stitching errors became noticeable when differences exceeded $\sim$15\%. Third, for steep curvature, peripheral cameras must image at increasingly oblique angles, causing both defocus and non-uniform magnification. For our system, the average DOF across cameras is 45~µm, resulting in a maximum curvature of 6.6~m$^{-1}$. These limitations collectively define the practical bounds of curvature that PANORAMA can accommodate using axial refocusing alone.

\begin{figure}[tbp!]
\centering
\includegraphics[width=\linewidth]{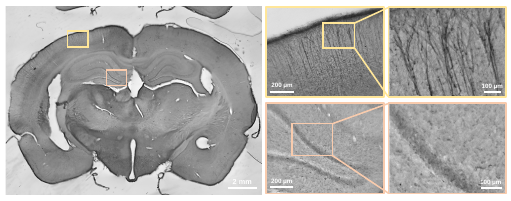}
\caption{Brightfield image of a rat brain slice acquired using a 48-camera reimaging system in one shot. Insets and zoom-ins show cellular-level structures such as dendrites (top right, yellow boxes) and hippocampal morphology (bottom right, pink boxes) over a large field of view.}
\label{fig3}
\end{figure}

We demonstrated the system’s large FOV and high resolution through brightfield imaging of a rat brain tissue slice. Benefit from the reimaging technique, the entire $\text{2~cm}^2$ section was captured in one shot, eliminating the need for lateral scanning and significantly reducing acquisition time. As shown in Fig.~\ref{fig3}, the stitched image from the 48-camera array covers 630~MP and resolves fine histological details, such as dendrites and the hippocampus, across the whole brain slice. The cellular-level structures and tissue morphology of large biospecimens that typically require sequential tiled images through mechanical scanning in the XY plane to cover the area, were visualized seamlessly across the panorama in one acquisition.

\begin{figure*}[t]
\centering
\includegraphics[width=\textwidth]{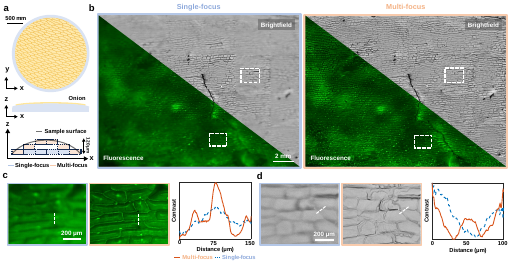}
\caption{Multi-focus snapshot imaging of curved biological samples using a tiled sub-camera array microscope. (a) Schematic of an onion epidermal sheet placed on a gently curved substrate. The system supports adaptive per-camera focus to match sample topology. (b) Whole-sample images acquired in both brightfield and fluorescence modes under single-focus (left) and multi-focus (right) configurations. The insects indicate the regions examined in detail. Fluorescence (c) and brightfield (d) zoom-ins and line profile respectively, under single-focus (left) versus multi-focus (middle).}
\label{fig4}
\end{figure*}

Moreover, the ability to refocus each sub-camera introduces a form of adaptive optics tailored to sample topology. Rather than being constrained to a flat focal plane, the microscope can accommodate samples with gentle curvature or uneven surfaces by pre-adjusting the focus locally. To illustrate the system’s capability to handle non-planar samples, we imaged onion epidermal cells ($\text{2.6~cm}^2$) placed on a gently curved surface in a single shot [Fig.~\ref{fig4}~(a)]. Both brightfield and fluorescence modes (with staining via fluorescein) were employed on the same sample by switching the illumination mode. {LED irradiance was set to 7 mW/cm² with 30 ms exposure for fluorescence imaging. The average fluorescence intensity of the signal region was 101.3, and the background mean was 10.4 with a standard deviation of 2.2, which results in a contrast-to-noise ratio of 42.3. We compared imaging the curved onion tissue under two configurations: one with a flat focal plane (all sub-cameras focused at the same depth), and another using a "curved" focal surface, achieved by independently adjusting each camera's focus.

With a flat focal setting, only part of a curved sample is in focus at a given plane, causing cells toward the periphery of the field to blur [Fig.~\ref{fig4}~(b)-\ref{fig4}~(d)]. In contrast, with per-camera focus adjustment the entire curved sheet of cells was captured sharply in one shot, yielding a uniformly high-resolution composite across the sample; by comparison, the flat-focus configuration showed noticeable defocus in regions away from the single best-focus plane [Fig.~\ref{fig4}~(b)-\ref{fig4}~(d)]. In brightfield, the stitched image clearly resolved the plant cell walls and overall cell arrangement over the curved region [Fig.~\ref{fig4}~(d)]. In the fluorescence channel, nuclei of the onion cells (stained using fluorescein) appeared as bright spots, corresponding to their locations within each cell [Fig.~\ref{fig4}~(c)]. This demonstrates that our proposed PANORAMA system, which tailors the focal plane to the specimen, preserves image sharpness across non-planar topographies. This capability can substantially streamline imaging of specimens that cannot lie perfectly flat, such as flexible tissues, curved membranes, or samples mounted on non-planar substrates. It negates the need for acquiring multiple focal stacks or physically flattening the sample, thus simplifying the workflow.

Looking ahead, the concept can be extended in several ways. Increasing the sensor count or using larger sensors could further expand the FOV, although at the cost of added system complexity. Integrating an automated focusing mechanism for each camera could enable rapid re-configuration of the focal surface to match different sample geometries without manual adjustment. The approach may also be combined with advanced computational imaging techniques; for instance, overlapping camera views could be exploited for 3D reconstruction or stereoscopic depth estimation, extending the system’s utility to volumetric imaging. In summary, the presented PANORAMA system provides a powerful proof-of-concept for snapshot gigapixel microscopy with microscopic resolution, offering significantly improved throughput and flexibility. By capturing wide-field, high-resolution images in a single shot and accommodating non-planar samples, this technique eliminates the need for mechanical scanning and complex focus stacking, potentially accelerating a range of biomedical and materials imaging applications.

\begin{backmatter}
\bmsection{Funding}
NIH Office of the Director (R44OD024879, R44OD036187); National Institute of Environmental Health Sciences (R44OD024879, R44OD036187); National Cancer Institute (R44CA285197, R44CA250877); National Institute of Biomedical Imaging and Bioengineering (R43EB030979); National Science Foundation (2036439); Duke-Coulter Translational Partnership Grant.

\bmsection{Disclosures}
MH: Ramona Optics Inc. (F, I, P, S). RH: Ramona Optics Inc. (F, I, P, S).

\bmsection{Data availability}
Data underlying the results presented in this paper are not publicly available at this time but may be obtained from the authors upon reasonable request.

\bmsection{Supplemental document}
See Supplement 1 for supporting content.
\end{backmatter}

\bibliography{main}

\bibliographyfullrefs{main}

\end{document}